\documentclass[review]{elsarticle}

\usepackage{lineno,hyperref}
\usepackage{graphicx,color}
\usepackage{amsmath}
\usepackage{amssymb}
\usepackage{color}

\modulolinenumbers[5]

\journal{Advances in Magnetism at the Joint European Magnetic Symposia 2018 (JEMS2018)}

\newcommand{\li}{Li$_2$FeSiO$_4$}
\newcommand{\ortho}{\textit{Pmn}2\(_1\)}

\newcommand{\mb}{\(\mu _{\rm B}\)}

\newcommand{\rk}[1]{\textcolor{black}{#1}}
\newcommand{\tn}{$T_{\rm N}$}
\newcommand{\lfs}{Li\(_2\)FeSiO\(_4\)}
\newcommand{\glfs}{$\gamma_{\rm II}$-Li\(_2\)FeSiO\(_4\)}

\begin{document}
\begin{frontmatter}
\title{Synthesis and Magnetism of a Li$_2$FeSiO$_4$ Single Crystal}

\author[KIP]{W.~Hergett\fnref{myfootnote}}\fntext[myfootnote]{Both authors contributed equally.}
\author[KIP]{M.~Jonak\fnref{myfootnote}}
\author[KIP]{J.~Werner}
\author[KIP]{F.~Billert}
\author[KIP]{S.~Sauerland}
\author[KIP]{C.~Koo}
\author[KIP]{C.~Neef}
\author[KIP,CAM]{R.~Klingeler\corref{mycorrespondingauthor}}
\cortext[mycorrespondingauthor]{Corresponding author}
\ead{klingeler@kip.uni-heidelberg.de}

\address[KIP]{Kirchhoff Institute of Physics, Heidelberg University, INF 227, D-69120 Heidelberg, Germany}
\address[CAM]{Centre for Advanced Materials, Heidelberg University, INF 225, D-69120 Heidelberg, Germany}


\begin{abstract}
A macroscopic single crystal of \glfs\ has been grown by means of the high-pressure \rk{optical} floating-zone technique. \rk{Static} magnetic susceptibility $\chi$ implies that the tetrahedrally-coordinated Fe$^{2+}$ ions are in the high-spin, $S=2$, state. While a sharp decrease in $\chi$ implies long-range antiferromagnetic order below \tn\ = 17.0(5)~K, the presence of a broad maximum at $T_{\rm m} = 28$~K suggests quasi-low-dimensional magnetism. Applying magnetic fields along the easy magnetic $a$-axis yields additional contributions to the susceptibility $\partial M/\partial B$ and magnetostriction for \rk{$B > 7$~T}, and an anomaly at $B\approx 14.8$~T.
\end{abstract}

\begin{keyword}
Single crystal growth; high-pressure \rk{optical} floating-zone technique; magnetisation; magnetostriction; low-dimensional magnetism;
\end{keyword}

\end{frontmatter}

\section{Introduction}

The orthosilicate \lfs\ is intensively studied as a new and high-capacity cathode material for lithium-ion batteries.~\cite{Islam2011} The orthorhombic $Pmnb$-structured polymorph \glfs\ exhibits tetrahedrally-coordinated Fe$^{\rm 2+}$ ions in a layered structure.~\cite{Lu2015,Nishimura2008,Boulineau2010} Here we present the magnetic characterization of a \glfs\ single crystal, grown by means of the high-pressure optical floating-zone method.~\cite{Neef}

\section{Experimental}


The crystal growth was carried out in a high-pressure floating-zone (FZ) furnace (HKZ, SciDre)~\cite{Neef,Wizent2014}. Polycrystalline \li\ starting materials used for the FZ process were synthesized by a conventional solid-state reaction. \rk{In particular}, a one-step synthesis method was applied to obtain \rk{the} \ortho\ polymorph, which was \rk{subsequently} reground. The reground powder was compacted into feed rods with diameters of 6~mm and typical lengths of 70 to 110~mm under an isostatic pressure of 60~MPa. \rk{The polycrystalline feed rods were used for the FZ-growth process to finally obtain single crystals of the title compound.} X-ray diffraction was performed in Bragg-Brentano geometry on a Bruker D8 Advance ECO diffractometer. A high-resolution X-Ray Laue camera (Photonic Science) was used to orient the single crystals, which were then cut into cuboids with respect to the crystallographic directions. Magnetisation in static magnetic fields up to 5~T was studied by means of a Quantum Design MPMS-XL5 SQUID magnetometer and in fields up to 15~T in a home-built vibrating sample magnetometer (VSM) \cite{vsm}. The field-induced length changes, $dL(B)/L$, were measured by means of a three-terminal high-resolution capacitance dilatometer operated in a variable temperature insert.~\cite{dilatometer} The magnetic field was applied along the direction of the measured length changes.

\section{Results and Discussion}

\begin{figure}
	\includegraphics[width=0.95\columnwidth,clip] {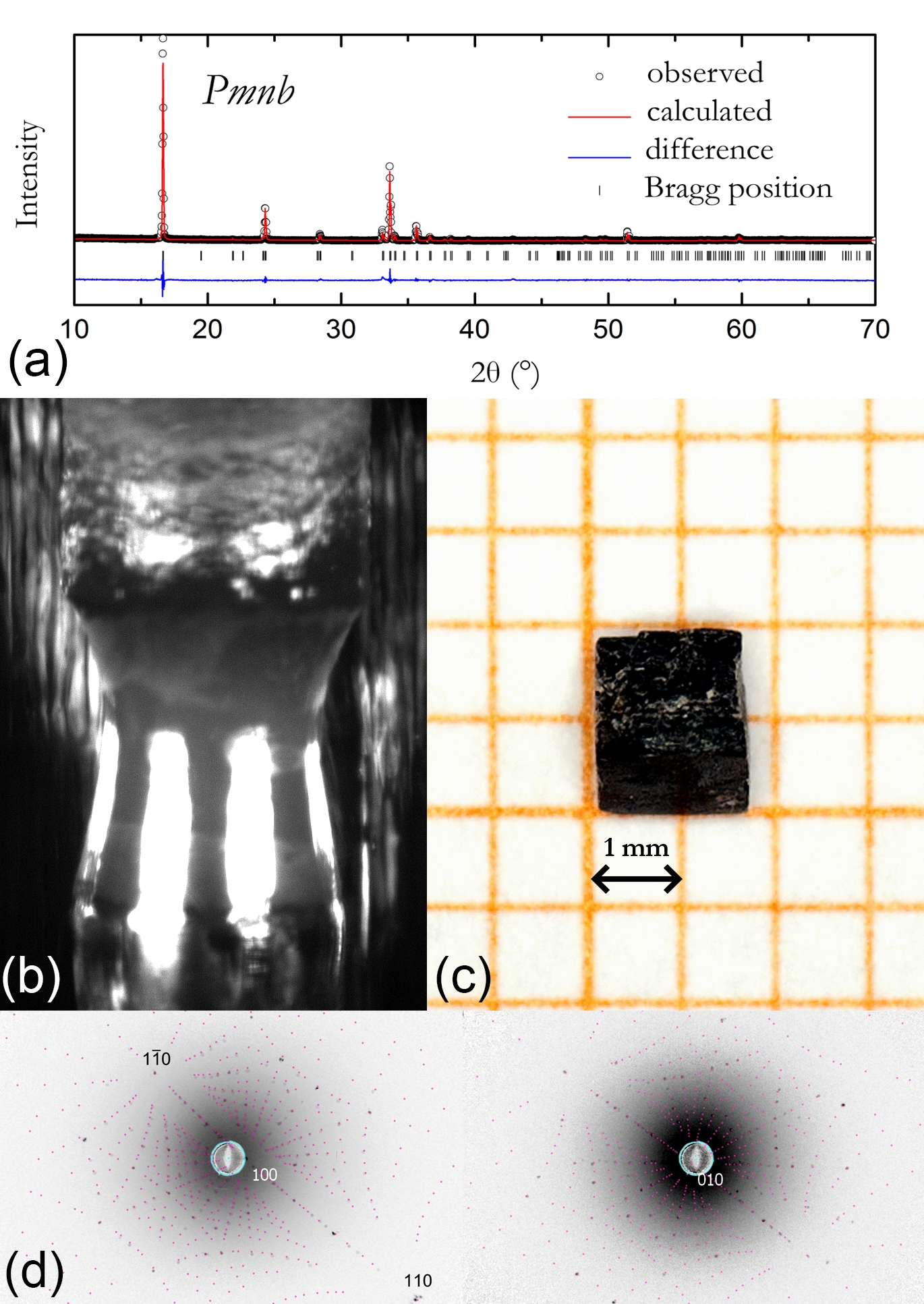}
	\caption{(a) XRD pattern collected from the powdered single crystal. (b) Photograph of the molten zone during the floating-zone growth of \li , and (c) resulting oriented single crystal. (d) Experimental (black spots) and simulated (purple spots) Laue patterns with the incident beam perpendicular to the \{100\} and \{010\} face of the crystal.}\label{xtal}
\end{figure}

\begin{figure}
\includegraphics[width=0.95\columnwidth,clip] {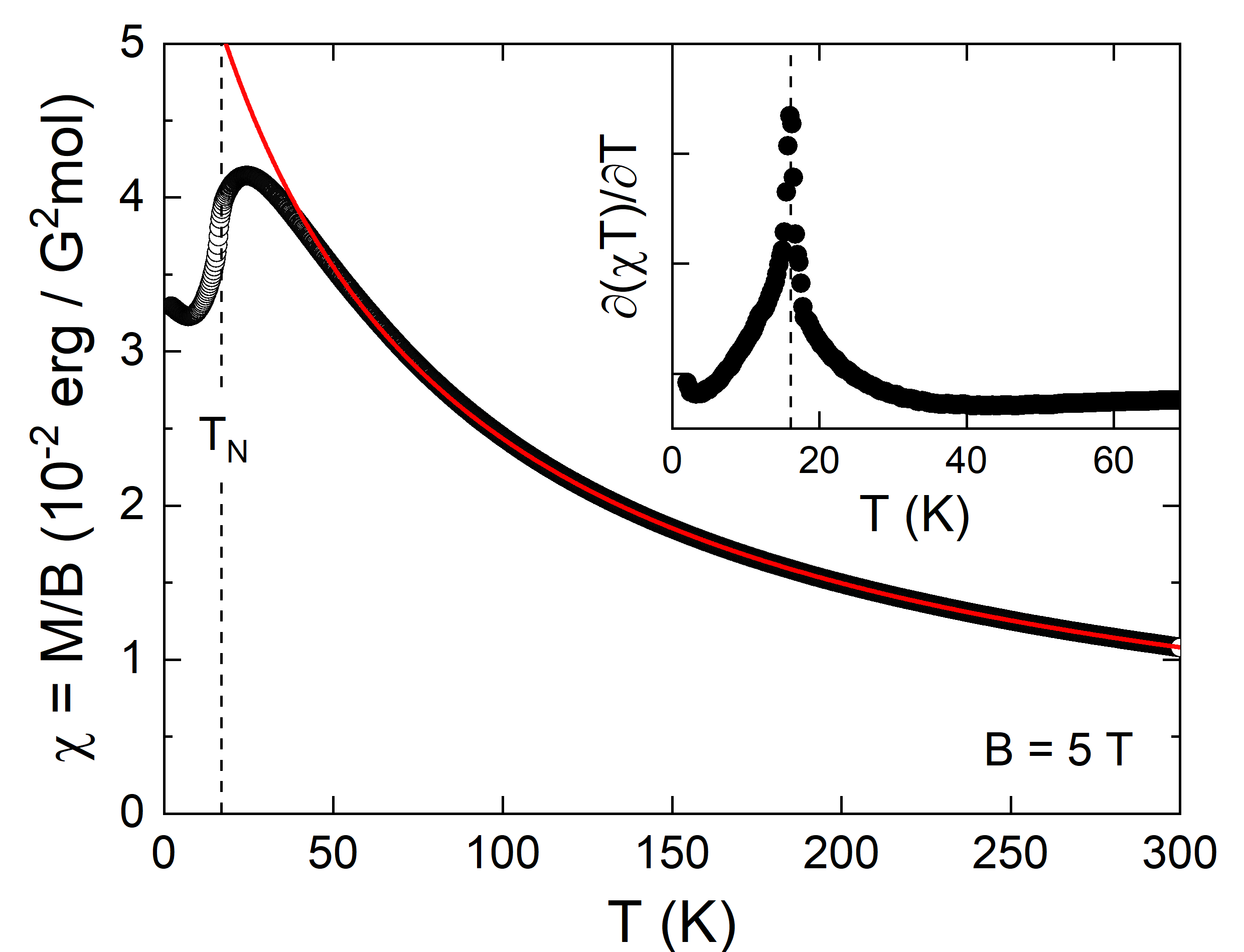}
\caption{Static \rk{magnetic} volume susceptibility $\chi = M/B$ of \li\ vs. temperature at $B=5$~T. The red line shows a fit by means of a Curie-Weiss-like model (see text). Inset: Derivative $\partial (\chi T)/\partial T$. The dashed vertical lines indicate \tn .}\label{chi}
\end{figure}

The high-pressure FZ-growth was carried out at a pulling rate of 10~mm/h, with feed and seed rods counter-rotating at 21 and 17~rpm, respectively. The growth was performed in a high-purity Ar atmosphere at a pressure of 30~bar and a flow rate of 0.02~$\rm{l/min}$. The temperature of the molten zone, measured in-situ by a two-colour pyrometer, was $\sim 1310^\circ$C.~\cite{Wizent} The elevated gas pressure of the growth atmosphere was applied to avoid Li$_2$O evaporation. These conditions were chosen \rk{in order} to form a stable molten zone as shown in Fig.~\ref{xtal}b. \rk{Despite utilising \ortho\ polycrystalline starting materials, the FZ-growth process yielded macroscopic single-crystalline samples of a $Pmnb$-polymorph of \li .} \rk{This is demonstrated by powder XRD on a ground piece of the single crystal depicted in Fig.~\ref{xtal}a. The obtained $Pmnb$-polymorph corresponds to the high-temperature phase of \li , while the low-temperature phase is $P12_1/n1$.}~\cite{Zhang2012} Laue diffractometry, as exemplified in Fig.~\ref{xtal}d, shows good crystallinity of the grown macroscopic grain. Fig.~\ref{xtal}c presents one of the oriented samples used for the thermodynamic measurements shown below. Note that we did not observe any influence of the growth rate, which was varied between 1.5 and 30~mm/h in our experiments, on the crystal structure of the obtained crystals. We therefore conclude that cooling rates are not a governing parameter.

\begin{figure}
\includegraphics[width=0.95\columnwidth,clip] {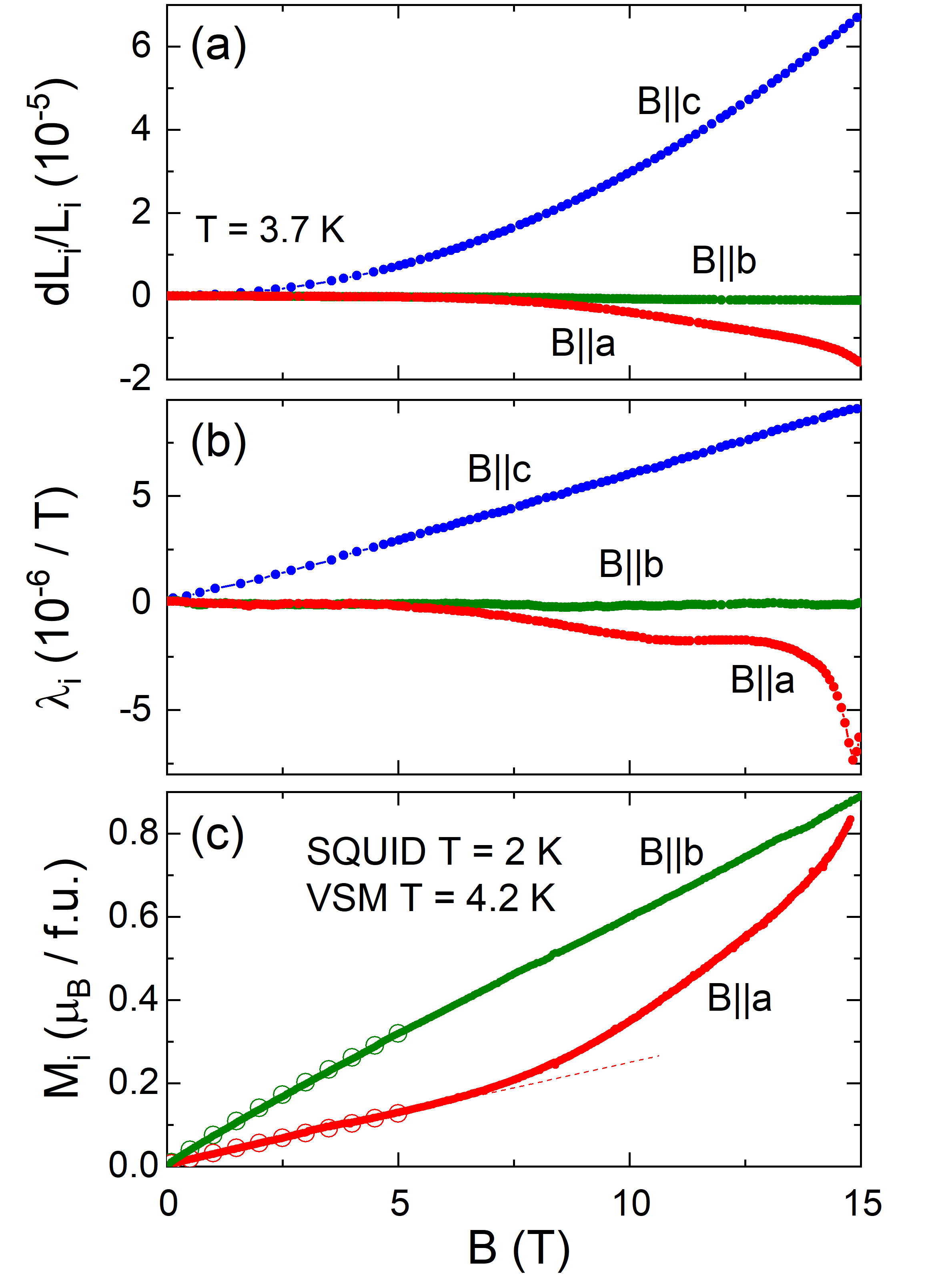}
\caption{(a) Magnetostriction $dL_i/L_i$ (at $T=3.7$~K), (b) corresponding magnetostriction coefficients $\lambda_{i}$, and (c) magnetisation $M_i$ (at $T=4.2$~K) of \li\ vs. external magnetic field for the different crystallographic axes $i=a,b,c$. Magnetisation data obtained in a VSM \rk{(solid symbols)} have been calibrated by SQUID data \rk{(open symbols)}. The dashed straight line is a guide to the eye.}\label{MMS}
\end{figure}

The static magnetic volume susceptibility $\chi = M/B$ of \li\ presented in Fig.~\ref{chi} reveals Curie-Weiss-like behaviour at high temperatures. At low temperatures, there is a broad maximum at around $T_{\rm m} = 28$~K, followed by a rapid decrease of $\chi$. A broad maximum in $\chi$ is characteristic of antiferromagnetic correlations and suggests quasi-low-dimensional magnetism. The observed sharp decrease in $\chi$ indicated by a distinct maximum in Fisher's specific heat $\partial (\chi T)/\partial T$ (inset of Fig.~\ref{chi}) implies long-range antiferromagnetic order at \tn\ = 17.0(5)~K. This value is similar to previous results on polycrystalline \li .~\cite{Zaghib2006b,Bini2013} Analysing the susceptibility in terms of a Curie-Weiss-like model $\chi = N_{\rm A}p^2\mu_{\rm B}^2/3k_{\rm B}(T+\Theta)+\chi_0$, with $N_{\rm A}$, \mb\ and $k_{\rm B}$ being Avogadro's number, the Bohr magneton, and Boltzmann's constant, yields the Weiss temperature $\Theta\approx 59(5)$~K and the Curie constant $C = 3.88$ ergK/(G\(^2\)mol). The \rk{positive} sign of $\Theta$ confirms predominant antiferromagnetic interactions. The value of $C$ implies the high-spin $S=2$-state of the Fe$^{2+}$ ions.

The effect of magnetic fields applied along the different crystallographic directions on the length and magnetisation of the oriented single crystal is shown in Fig~\ref{MMS}. Fields $B<7$~T parallel to the easy magnetic $a$-axis do not cause significant magnetostriction. However, further increase of $B$ results in a sizeable shrinking of the $a$-axis, and there is an indication of a phase transition slightly below the highest field accessible in our experiment, i.e. at around 14.8~T. Field-induced length changes are associated with additional contributions to the magnetic susceptibility $\partial M/\partial B$ as shown in Fig.~\ref{MMS}c. To be specific, the linear slope in $M$ vs. $B$ develops a left-bending character at $B \gtrsim 7$~T. In contrast, there are no particular features for $B\perp a$-axis. As discussed above, the observed features in $M$ vs. $B$ and $\lambda$ at the maximum accessible field may indicate the presence of a phase transition. For example, the data would agree with a field-driven spin re-orientation. Further studies at higher magnetic fields are needed to clarify this issue.

\section{Summary}

We present the first growth of a macroscopic \glfs\ single crystal by means of the high-pressure optical floating-zone technique. Our data show high quality of the grown crystal. While the magnetic susceptibility confirms evolution of long-range antiferromagnetic order at \tn\ = 17.0(5)~K, we also find evidence of quasi-low-dimensional magnetism. Applying magnetic field \rk{$B > 7$~T} along the easy magnetic $a$-axis yields additional contributions to the magnetic susceptibility $\partial M/\partial B$ and to the magnetostriction, as well as an anomaly indicative of a field-induced phase transition at $B\approx 14.8$~T.

\section{Acknowledgements}
The authors thank H.~Wadepohl and H.~P.~Meyer for valuable experimental support. J.~W. acknowledges support from the HGSFP. W.~H. and R.~K. acknowledge support by the BMBF German-Egypt Research Fund GERF IV via project 01DH17036. M.~J. acknowledges funding via the LGF-Promotionskolleg \textit{Basic Building Blocks for Quantum Enabled Technologies}.


\end{document}